\documentclass[twocolumn,aps,pra,showpacs]{revtex4}

\begin{document}
\title{How much larger quantum correlations are than classical ones}
\author{Ad\'{a}n Cabello}
\email{adan@us.es}
\affiliation{Departamento de F\'{\i}sica Aplicada II,
Universidad de Sevilla, 41012 Sevilla, Spain}
\date{\today}


\begin{abstract}
Considering as distance between two two-party correlations the
minimum number of half local results one party must toggle in
order to turn one correlation into the other, we show that the
volume of the set of physically obtainable correlations in the
Einstein-Podolsky-Rosen-Bell scenario is $(3 \pi/8)^2 \approx
1.388$ larger than the volume of the set of correlations
obtainable in local deterministic or probabilistic hidden-variable
theories, but is only $3 \pi^2/32 \approx 0.925$ of the volume
allowed by arbitrary causal (i.e., no-signaling) theories.
\end{abstract}


\pacs{03.65.Ud,
03.65.Ta}
\maketitle


\section{Introduction}


Quantum information (that is, information carried by microscopic
systems described by quantum mechanics such as atoms or photons)
can connect two spacelike separated observers by correlations that
cannot be explained by classical communication. This fact,
revealed by Bell's inequalities and violations
thereof~\cite{Bell64,CHSH69,CH74,Bell76}, is behind common
statements such as that quantum correlations are ``stronger'' or
``larger'' than classical ones, or that quantum-mechanical systems
may be ``further correlated'' than those obeying classical
physics~\cite{Vedral02}, and has been described as ``the most
profound discovery of science''~\cite{Stapp75}. Given its
fundamental importance, it is surprising that the question of how
much ``larger'' than classical correlations quantum correlations
are has not, to my knowledge, a precise answer beyond the fact
that quantum mechanics violates Clauser-Horne-Shimony-Holt (CHSH)
inequalities~\cite{CHSH69} up to $2 \sqrt{2}$ (Tsirelson's
bound~\cite{Tsirelson80}), while the bound for local
hidden-variable theories is just~$2$~\cite{CHSH69}. To be more
specific, if by ${\cal Q}$ we denote the set of all correlations
allowed by quantum mechanics in a given experimental scenario, and
by ${\cal C}$ the corresponding set of correlations allowed by any
local deterministic~\cite{Bell64,CHSH69} or
probabilistic~\cite{CH74,Bell76} hidden-variable theory (the set
of correlations allowed by both types of theories turns out to be
the same~\cite{CH74}), a more precise measure of how much larger
quantum correlations are when compared to classical ones would be
the ratio between the volumes (i.e., the hyper-volumes or
contents) of both sets, $V_{\cal Q}/V_{\cal C}$. To my knowledge,
the value $V_{\cal Q}/V_{\cal C}$, even for the simplest
nontrivial experimental scenario, cannot be found anywhere in the
literature.

Moreover, another interesting problem is why quantum correlations
cannot be even ``larger'' than they are. For instance, Popescu and
Rohrlich raised the question of whether the no-signaling condition
could restrict the set of physically obtainable correlations to
those predicted by quantum mechanics~\cite{PR94}. Although they
proved this conjecture to be false~\cite{PR94} (however,
see~\cite{BM04}), a precise measurement of the ratio between the
volume of quantum correlations and the volume of the set ${\cal
L}$ of all possible correlations allowed by any arbitrary causal
(i.e., no-signaling) theory, $V_{\cal Q}/V_{\cal L}$, still cannot
be found in the literature.


\subsection{\label{natural}A natural distance in the space of correlations}


The three spaces compared in this paper have quite different
status:

${\cal Q}$, the set of correlations obtainable by quantum
mechanics, gives a perfect account of all preparations and
measurements which are possible in nature.

${\cal C}$, the set of correlations obtainable by local
deterministic or probabilistic hidden-variable theories, does not
give a complete account of all possible preparations and
measurements.

${\cal L}$, the set of correlations obtainable by any arbitrary
causal (i.e., no-signaling) theory, is nonphysical, in the sense
that it admits preparations and measurements that are impossible.

The volume of a space of correlations depends on the choice of a
measure. In order to compare these three sets, the first problem
to address is finding a common measure with a clear meaning for
the three sets. Some criteria proposed in the literature are:

(i) The minimum amount of classical communication between the
parties necessary to reproduce some correlations, assuming that
the parties share correlations belonging to ${\cal
C}$~\cite{Pironio03}.

(ii) The number of trials of the experiment required to observe
that a set of correlations provides a substantial violation of the
predictions of ${\cal C}$~\cite{VGG03}.

(iii) The reduction of the amount of classical communication
needed for some specific distributed computational task compared
to the classical communication needed when assuming correlations
belonging to ${\cal C}$~\cite{BLMPPR05}.

Each of these proposals would lead to a different distance in the
space of correlations, and all of these distances have their own
problems. All of them privilege some or all the elements of ${\cal
C}$. In particular, (i) and (ii) would assign a volume zero to
${\cal C}$, while (iii) would introduce a distance which would be
sometimes negative.

In this paper we shall define as distance between two two-party
correlations $\langle A_i B_j \rangle$ and $\langle A_k B_l
\rangle$ the minimum number of half local results one party must
toggle in order to turn one correlation into the other. For
instance, suppose that the results of repeating Alice's experiment
$A_i$ and Bob's experiment $B_j$ a million times are completely
uncorrelated, i.e., $\langle A_i B_j \rangle = 0$. A way of
increasing the correlation is that Alice toggles some of her
results ($-1$ or $+1$) in order to be perfectly correlated with
Bob's corresponding results. If they want to increase the
correlation to $0.5$, Alice must toggle at least $250,000$ results
or, equivalently, $0.25$ results per single experiment.
Analogously, to change from $\langle A_i B_j \rangle = 0.7$ to
$0.1$, Alice must toggle a minimum of $0.3$ results per
experiment. As can be easily seen, this distance induces a uniform
measure over the space of correlations.

Besides its clear physical meaning, this measure has some other
advantages. It only privileges the point where the results of the
experiments of the two parties are completely uncorrelated, what
is reasonable from a physical perspective. Moreover, the
corresponding volumes have a simple interpretation: the
probability that four random numbers between $-1$ and $+1$ belong
to ${\cal C}$ (i.e., satisfy the CHSH inequalities) is the volume
of ${\cal C}$ divided by the volume of ${\cal L}$, etc.
Consequently, $V_{\cal Q}/V_{\cal C}$ is a reasonable measure of
how much larger ${\cal Q}$ is than ${\cal C}$, and $V_{\cal
Q}/V_{\cal L}$ is a physically reasonable measure of how much
larger the set of correlations could be without violating the
no-signaling condition.


\subsection{The Einstein-Podolsky-Rosen (EPR)-Bell scenario}


The EPR-Bell
scenario~\cite{EPR35,Bohm51,Bell64,CHSH69,CH74,Bell76} is the
simplest and most basic one where the difference between classical
and quantum correlations arises. It consists of two alternative
dichotomic experiments (i.e., having only two possible outcomes,
which we can label $\pm 1$), $A_0$ or $A_1$, on a subsystem $A$,
and another two alternative dichotomic experiments, $B_0$ or
$B_1$, on a distant subsystem $B$. In this scenario the set of
correlations $\langle A_i B_j \rangle$ is four-dimensional. It is
the standard scenario in which the Einstein-Podolsky-Rosen (EPR)
argument~\cite{EPR35} is presented~\cite{Bohm51} and in which
violations of Bell's inequalities~\cite{Bell64,CHSH69,CH74,Bell76}
have been experimentally
demonstrated~\cite{ADR82,WJSWZ98,Aspect99}, and also the one in
which recent experiments to search for stronger-than-quantum
correlations~\cite{FS04a,Cabello04,BCCDRR04,TSBKCW04} has been
discussed. On the other hand, the EPR-Bell scenario is contained
in any experimental scenario involving more subsystems, more
experiments per subsystem, or more discrete outcomes per
experiment.

Hereafter, we shall consider the EPR-Bell scenario and will not
make any assumptions about the type of physical subsystems
considered, their state before the local experiments, or the type
of local experiments performed.

The essential elements of the EPR-Bell scenario may be summarized
as two boxes, each with two possible inputs (the two alternative
local experiments) and two possible outcomes~\cite{Mermin81}. Each
possible pair of boxes can be characterized by a set of $2^4$
joint probabilities for the various possible outcomes: $P(A_i=a,
B_j=b)$, $\{a,b\} \in \{-1,1\}$. These probabilities satisfy
positivity [i.e., for all $A_i$, $B_j$, and $\{a,b\} \in
\{-1,1\}$, $P(A_i=a, B_j=b) \ge 0$], and a normalization condition
[i.e., for all $A_i$, $B_j$, and $\{a,b\} \in \{-1,1\}$,
$\sum_{\{a,b\} \in \{-1,1\}} P(A_i=a, B_j=b) = 1$], and constitute
(without taking into account further constraints) a set of
dimension~$12$ (of dimension~$8$ if we impose the no-signaling
constraint). The set of correlations is a four-dimensional
projection of the set of joint probabilities. The connection
between both sets is given by
\begin{equation}
\langle A_i B_j \rangle = \sum_{k,l \in \{-1,1\}} a_k b_l
P(A_i=a_k,B_j=b_l).
\end{equation}

For the EPR-Bell scenario, and assuming the uniform measure
induced by the distance introduced in Sec.~\ref{natural}, in
Sec.~\ref{II} we calculate $V_{\cal C}$, in Sec.~\ref{III} we
calculate $V_{\cal L}$, and in Sec.~\ref{IV} we calculate $V_{\cal
Q}$ and both $V_{\cal Q}/V_{\cal C}$ and $V_{\cal Q}/V_{\cal L}$,
and quantify the success of several previous attempts to
characterize ${\cal Q}$ by means of linear~\cite{Tsirelson80} and
quadratic~\cite{Uffink02} Bell-like inequalities. Finally, in
Sec.~\ref{V} we suggest further lines of research.


\section{\label{II} Correlations allowed by local hidden-variable
theories}


Froissart~\cite{Froissart81} and Fine~\cite{Fine82a,Fine82b} (see
also~\cite{Pitowsky86,Pitowsky89}) proved that, for the EPR-Bell
scenario, the set of all joint probabilities attainable by local
hidden-variable theories (i.e., theories in which the local
variables determine the probability distribution for the different
possible results of the local experiments) is an eight-dimensional
polytope with 16~vertices and 24~faces. The four-dimensional
projection corresponding to the set ${\cal C}$ of all correlations
that can be attained by local hidden-variable theories is defined
by eight CHSH inequalities. To be precise, a set of four real
numbers $\langle A_i B_j\rangle$ ($i,j=0,1$) belongs to ${\cal
C}$, i.e., is attainable by local hidden-variable theories, if and
only if
\begin{equation}
|\langle A_0 B_0\rangle + \langle A_0 B_1\rangle + \langle A_1
B_0\rangle + \langle A_1 B_1\rangle - 2 \langle A_i B_j\rangle |
\le 2,
\label{Fine}
\end{equation}
for all $i,j=0,1$. The volume of this four-dimensional set ${\cal
C}$ can be easily calculated,
\begin{equation}
V_{\cal C}= {2^5 \over 3}.
\label{VC}
\end{equation}


\section{\label{III}Correlations allowed by arbitrary causal theories}


Let us consider theories where the only restriction is that
signaling is forbidden (i.e., the two distant observers cannot
signal to one another via their choice of input). The no-signaling
condition restricts the set of joint probabilities. The
no-signaling condition imposes that the marginal probabilities
$P(B_j=b)$ [$P(A_i=a)$] should be independent of the choice of
$A_i$ [$B_j$], for all $B_j$ and $b \in \{-1,1\}$ [for all $A_i$
and $a \in \{-1,1\}$]. This implies eight restrictions on the set
of joint probabilities: $P(A_0=1,B_j=b)+P(A_0=-1,B_j=b) =
P(A_1=1,B_j=b)+P(A_1=-1,B_j=b)$ and
$P(A_i=a,B_0=1)+P(A_i=a,B_0=-1) = P(A_i=a,B_1=1)+P(A_i=a,B_1=-1)$
for all $A_i$, $B_j$, and $\{a,b\} \in \{-1,1\}$, so that the set
of all possible joint probabilities satisfying the no-signaling
condition has dimension~$8$. This set is a convex polytope with
24~vertices and 16~faces~\cite{Tsirelson93}. Most of the points of
this set are not physically realizable; however, its potential
usage as an information theoretic resource has been recently
investigated~\cite{BLMPPR05}.

However, the restrictions imposed on the set of joint
probabilities by the no-signaling condition do not imply new
nontrivial restrictions on the set of correlations $\langle A_i
B_j\rangle$. There are either sets of joint probabilities
violating no-signaling but satisfying inequalities (\ref{Fine})
[for instance, $P(A_0=1,B_0=1)=P(A_0=1,B_0=-1)=
P(A_0=-1,B_1=1)=P(A_0=-1,B_1=-1)= P(A_1=1,B_0=1)=P(A_1=1,B_0=-1)=
P(A_1=-1,B_1=1)=P(A_1=-1,B_1=-1)=1/2$)], and sets satisfying
no-signaling but maximally violating (\ref{Fine}) [for instance,
$P(A_0=1,B_0=1)=P(A_0=-1,B_0=-1)= P(A_0=1,B_1=1)=P(A_0=-1,B_1=-1)=
P(A_1=1,B_0=1)=P(A_1=-1,B_0=-1)=
P(A_1=1,B_1=-1)=P(A_1=-1,B_1=1)=1/2$ \cite{PR94}]. Therefore the
set ${\cal L}$ of all correlations that can be attained by
arbitrary causal theories is simply defined by the
eight~inequalities
\begin{equation}
|\langle A_i B_j\rangle| \le 1,
\label{corrdef}
\end{equation}
for $i,j=0,1$. ${\cal L}$ is a four-dimensional cube (a
tessaract). Its volume is
\begin{equation}
V_{\cal L}= {2^4}.
\label{VL}
\end{equation}
Comparing Eqs. (\ref{VC}) and (\ref{VL}), we conclude that the
volume of the set of correlations attainable by local
hidden-variable theories is just $2/3$ of that allowed by
arbitrary causal theories.


\section{\label{IV}Correlations allowed by quantum mechanics}


Although the necessary and sufficient conditions defining the set
of quantum correlations ${\cal Q}$ in the EPR-Bell scenario have
long been known~\cite{Tsirelson85}, surprisingly they are rarely
mentioned in the literature. In contrast, some necessary (but not
sufficient) conditions for a set of four correlations to belong to
${\cal Q}$ are much better known. Let us start by reviewing the
two most famous ones.


\subsection{Tsirelson's linear inequalities}


Tsirelson~\cite{Tsirelson80} showed that, for {\em any} quantum
state $\rho$, the four quantum correlations $\langle A_i
B_j\rangle$, for $i,j=0,1$, must satisfy eight linear inequalities
(usually referred to as Tsirelson's inequalities, but here we
shall call them Tsirelson's {\em linear} inequalities), which can
be written as
\begin{equation}
|\langle A_0 B_0\rangle + \langle A_0 B_1\rangle + \langle A_1
B_0\rangle + \langle A_1 B_1\rangle - 2 \langle A_i B_j\rangle |
\le 2 \sqrt{2},
\label{Tsirelsonbound}
\end{equation}
for all $i,j=0,1$~\cite{Tsirelson93}. Different proofs of the
inequalities~(\ref{Tsirelsonbound}) can be found
in~\cite{KT85,Landau87,KT92}. Quantum mechanics predicts
violations of the CHSH inequalities~(\ref{Fine}) up to $2
\sqrt{2}$. Such violations can be obtained with pure~\cite{CHSH69}
or mixed states~\cite{BMR92}. A method for deriving maximal
violations of Bell-type inequalities can be found in~\cite{FS04b}.

The volume of the set ${\cal T}$ defined by Tsirelson's linear
inequalities~(\ref{Tsirelsonbound}) is
\begin{equation}
V_{\cal T} \approx 0.961 \times 2^4.
\label{VT}
\end{equation}


\subsection{Uffink's quadratic inequalities}


Uffink's quadratic inequalities~\cite{Uffink02} provide a more
restrictive necessary (but still not sufficient) condition for the
correlations to be attainable by quantum mechanics. According to
Uffink, the four correlations must satisfy the following
inequalities:
\begin{eqnarray}
(\langle A_0 B_0\rangle + \langle A_1 B_1\rangle)^2 + (\langle A_0
B_1\rangle - \langle A_1 B_0\rangle)^2 \le 4, \nonumber \\
(\langle A_0 B_0\rangle - \langle A_1 B_1\rangle)^2 + (\langle A_0
B_1\rangle + \langle A_1 B_0\rangle)^2 \le 4
\label{Uffink}.
\end{eqnarray}

The volume of the set ${\cal U}$ defined by Uffink's quadratic
inequalities~(\ref{Uffink}) is
\begin{equation}
V_{\cal U} \approx 0.950 \times 2^4.
\label{VU}
\end{equation}


\subsection{Tsirelson's and Landau's inequalities}


However, both Tsirelson's linear
inequalities~(\ref{Tsirelsonbound}) and Uffink's quadratic
inequalities (\ref{Uffink}) are necessary {\em but not sufficient}
conditions for the correlations to be attainable by local
measurements on subsystems of a composite physical system prepared
in a quantum state. Although rarely mentioned in the literature,
to my knowledge, there are three equivalent sets of necessary and
sufficient conditions to define the set ${\cal Q}$ of correlations
attainable by quantum mechanics. The first was provided by
Tsirelson~\cite{Tsirelson85}. According to Tsirelson, a set of
four correlations $\langle A_i B_j\rangle$ ($i,j=0,1$) is
realizable in quantum mechanics (i.e., belongs to ${\cal Q}$) if
at least one of the following two inequalities holds:
\begin{eqnarray}
0 & \le & (\langle A_0 B_1\rangle \langle A_1 B_0\rangle-\langle
A_0
B_0\rangle \langle A_1 B_1\rangle) \nonumber \\
& & \times (\langle A_0 B_0\rangle \langle A_0 B_1\rangle-\langle
A_1
B_0\rangle \langle A_1 B_1\rangle) \nonumber \\
& & \times (\langle A_0 B_0\rangle \langle A_1 B_0\rangle-\langle
A_0
B_1\rangle \langle A_1 B_1\rangle) \nonumber \\
& \le & {1 \over 4} \left(\sum_{i,j} \langle A_i B_j\rangle^2
\right)^2 -{1 \over 2} \sum_{i,j} \langle A_i B_j\rangle^4
\nonumber \\ & & -2 \prod_{i,j} \langle A_i B_j\rangle,
\label{tsi01} \\
0 & \le & 2 \max_{i,j} \langle A_i B_j\rangle^4-(\max_{i,j}
\langle A_i B_j\rangle^2) \left(\sum_{i,j} \langle A_i
B_j\rangle^2\right) \nonumber \\ & & + 2 \prod_{i,j} \langle A_i
B_j\rangle. \label{tsi02}
\end{eqnarray}

The second characterization of ${\cal Q}$ is due to
Landau~\cite{Landau88}. According to him, four correlations belong
to ${\cal Q}$ if and only if they satisfy the following
inequalities:
\begin{eqnarray}
|\langle A_0 B_0\rangle \langle A_0 B_1\rangle - \langle A_1
B_0\rangle \langle A_1 B_1\rangle| \nonumber \\ \le
\sqrt{1-\langle A_0 B_0\rangle^2} \sqrt{1-\langle A_0
B_1\rangle^2}
\nonumber \\
+ \sqrt{1-\langle A_1 B_0\rangle^2} \sqrt{1-\langle A_1
B_1\rangle^2}.
\label{landau}
\end{eqnarray}
These inequalities (\ref{landau}) are equivalent to inequalities
(\ref{tsi01}) and (\ref{tsi02})~\cite{Tsirelson93}.

The third equivalent definition of ${\cal Q}$ can be explicitly
found for the first time in~\cite{Tsirelson93} (although it can be
easily derived from the results in~\cite{Landau88}). According to
this, four correlations belong to ${\cal Q}$ if and only if they
satisfy the following eight inequalities:
\begin{eqnarray}
|\arcsin{\langle A_0 B_0\rangle} + \arcsin{\langle A_0 B_1\rangle}
+ \arcsin{\langle A_1 B_0\rangle} \nonumber \\
+ \arcsin{\langle A_1 B_1\rangle} - 2 \arcsin{\langle A_i
B_j\rangle}| \le \pi,
\label{Landautsirelson}
\end{eqnarray}
for all $i,j=0,1$. Using inequalities~(\ref{Landautsirelson}) to
describe ${\cal Q}$ has the advantage of being analogous to using
inequalities~(\ref{Fine}) to describe ${\cal C}$. These
inequalities~(\ref{Landautsirelson}) have been recently
rediscovered by Masanes~\cite{Masanes03} (see also~\cite{WW01}).


\subsection{Main results}


The simplest way to calculate the volume of ${\cal Q}$, which is a
four-dimensional convex set~\cite{Tsirelson93}, is by using
expression~(\ref{landau}). Then, it can be seen that
\begin{equation}
V_{\cal Q}= {3 \pi^2 \over 2} \approx 0.925 \times 2^4.
\label{VQ}
\end{equation}
Therefore the ratio between the volumes of the set of quantum
correlations and those allowed by local hidden-variable theories,
which, as explained in Sec.~\ref{natural}, is a natural measure of
how much larger than classical correlations quantum correlations
are for the EPR-Bell scenario, is
\begin{equation}
{V_{\cal Q} \over V_{\cal C}} = \left({3 \pi \over 8}\right)^2
\approx 1.388.
\end{equation}
Although the number $\pi$ is ubiquitous in
nature~\cite{BBB99,ELW99}, it is very surprising to find it again
in this context.

On the other hand, the ratio between the volumes of the set of
quantum correlations and those allowed by arbitrary causal
theories is
\begin{equation}
{V_{\cal Q} \over V_{\cal L}} = {3 \pi^2 \over 32} \approx 0.925.
\label{VQVL}
\end{equation}
Popescu and Rohrlich's question was why quantum correlations do
not violate the CHSH-Bell inequalities ``more'' than they
do~\cite{PR94}. Result (\ref{VQVL}) allows us to quantify how much
larger than the set of quantum correlations the set of possible
correlations could be: $7.5$\% of the, in principle, possible sets
of four correlations never occur in nature.

In addition, we can use $V_{\cal Q}$ to quantify the success of
previous characterizations of ${\cal Q}$. For instance, comparing
$V_{\cal T}$, given by Eq.~(\ref{VT}), with $V_{\cal Q}$, given by
Eq.~(\ref{VQ}), we obtain that the set ${\cal T}$ defined by
Tsirelson's lineal inequalities (\ref{Tsirelsonbound}) is $3.8$\%
larger than ${\cal Q}$ (i.e., $3.7$\% of the sets of four
correlations belonging to ${\cal T}$ are not actually achievable
by quantum mechanics). On the other hand, comparing $V_{\cal U}$,
given by Eq.~(\ref{VU}), with $V_{\cal Q}$, we obtain that the set
${\cal U}$ defined by Uffink's lineal inequalities (\ref{Uffink})
is $2.6$\% larger than ${\cal Q}$ (i.e., $2.6$\% of the sets of
four correlations belonging to ${\cal T}$ are not actually
achievable by quantum mechanics).


\section{\label{V}Further lines of research}


In this paper we have investigated the basic scenario where
quantum correlations are different than local ones: $N=2$ parties,
each of them choosing between $M=2$ alternative local experiments,
each of which have $D=2$ possible outcomes. There are two basic
lines to extend these results:

One line for future research is to investigate how the ratios
$V_{\cal Q}/V_{\cal C}$ and $V_{\cal Q}/V_{\cal L}$ evolve in more
complex scenarios, namely those involving more parties, more
alternative local experiments per party, and more outcomes per
experiment. In particular, it would be interesting to know how
$V_{\cal Q}/V_{\cal C}$ evolves with $N$, assuming that $M=2$ and
$D=2$ are fixed. For this case, Mermin showed a Bell inequality
which is violated by quantum predictions by an amount that grows
exponentially with $N$~\cite{Mermin90}. Does this happen with
$V_{\cal Q}/V_{\cal C}$ or, on the contrary, does $V_{\cal
Q}/V_{\cal C}$ decrease with $N$ and is there a ``classical
limit''? The analytical expression of $V_{\cal Q}/V_{\cal C}$ as a
function of $N$ would require us to know what are the necessary
and sufficient conditions which define the set of quantum
correlations for these more complex scenarios, something which is
still an open problem. However, some numerical research can be
attempted.

On the other hand, we have just paid attention to the space of
{\em correlations}, which is just a four-dimensional projection of
the (eight-dimensional, if we assume no-signaling) space of {\em
joint probabilities}. In general, there are many different sets of
joint probabilities giving the same set of correlations. For
instance, in the EPR-Bell scenario, the fact that four numbers
satisfy Eq.~(\ref{Landautsirelson}) only implies that there exists
at least one compatible set of eight joint probability
distributions obtainable by performing measurements on a quantum
state, but it does not mean that {\em any} possible set of eight
joint probabilities satisfying Eq.~(\ref{Landautsirelson}) is
obtainable by performing measurements on a quantum state.
Therefore another interesting line for future research would be
calculating the volume of the set of quantum joint probabilities
and comparing it with the volume of the joint probabilities
allowed by local hidden-variable theories. The former could be
calculated numerically, since the necessary and sufficient
condition for eight probabilities to be allowed by quantum
mechanics can be expressed as a ``Russian doll''-type set of
theorems~\cite{Tsirelson80} that can be implemented as a computer
program. However, the analytical solution could be much more
difficult to obtain, since it still is an open problem whether or
not it is possible to define the set of quantum joint
probabilities by means of a set of polynomial, or even analytical,
inequalities~\cite{Tsirelson93}.


\section{Conclusions}


How much ``larger'' than classical correlations are quantum
correlations? We have shown that the volume of the set of
physically obtainable correlations $\langle A_i B_j \rangle$ in
the EPR-Bell scenario, namely where $A_0$ and $A_1$ ($B_0$ and
$B_1$) are two alternative dichotomic experiments on subsystem $A$
(on a distant subsystem $B$), assumed to be those obtainable by
local measurements on quantum states, is $(3 \pi/8)^2 \approx
1.388$~larger than the volume of the set of correlations
obtainable by local hidden-variable theories.

How much ``larger'' could, in principle, the set of correlations
be? We have shown that the set of quantum correlations is only $3
\pi^2/32 \approx 0.925$~of the volume allowed by arbitrary causal
(i.e., no-signaling) theories.

In other words, given four~real random numbers between $-1$ and
$+1$, the probability for them to be reproducible by local
hidden-variable theories is~$2/3$. The probability for them to be
physically obtainable (i.e., reproducible by quantum mechanics) is
$3 \pi^2/32 \approx 0.925$. The probability for them to be
reachable by arbitrary causal theories is~$1$.

In addition, we have shown that the sets defined by Tsirelson's
linear inequalities~\cite{Tsirelson80} and Uffink's quadratic
inequalities~\cite{Uffink02} contain $3.7$\% and $2.6$\%,
respectively, of elements that cannot be obtained within quantum
mechanics.


\section*{Acknowledgments}


We thank L.~Masanes, B.~S.~Tsirelson, and R.~F.~Werner for helpful
discussions and comments, I.~Pitowsky for references, and the
Spanish Ministerio de Ciencia y Tecnolog\'{\i}a Grant
No.~BFM2002-02815 and the Junta de Andaluc\'{\i}a Grant
No.~FQM-239 for support.



\end{document}